\documentclass[12pt]{article}

\usepackage{ifthen,pifont,natbib,comment,graphicx}
\usepackage{mathrsfs,amssymb}
\usepackage{times,avant}
\usepackage{xy}\xyoption{all} \xyoption{poly} \xyoption{knot}
\usepackage{amsfonts}
\usepackage{amsmath,amssymb}
\usepackage{enumerate}
\usepackage{multirow}
\usepackage[figuresright]{rotating}
\input xy
\xyoption{all}
\usepackage{epstopdf}
\usepackage{color}

\newcommand{\bOmega}{\mbox{\boldmath $\Omega$}}
\newcommand{\brho}{\mbox{\boldmath $\rho$}}
\newcommand{\bet}{\mbox{\boldmath $\beta$}}

\newcommand{\bGamma}{\mbox{\boldmath $\Gamma$}}
\newcommand{\bgamma}{\mbox{\boldmath $\gamma$}}
\newcommand{\blambda}{\mbox{\boldmath $\lambda$}}

\newcommand{\bLambda}{\mbox{\boldmath $\Lambda$}}

\newcommand{\bmu}{\mbox{\boldmath $\mu$}}

\usepackage[figuresright]{rotating}

\begin{document}


\title{\bf Bayesian inference on dependence in multivariate longitudinal data }

\author{
Hongxia Yang$^{*a}$, Fan Li$^{b}$, Enrique F. Schisterman$^{c}$,\\
Sunni L. Mumford$^{c}$   and David Dunson$^{b}$} \vspace{4in}
\date{}

\maketitle

\vspace{3in}

\small{ \noindent $^a$Watson Research Center (Yorktown), IBM,
Statistical Analysis \&
Forecasting, Mathematical Sciences Department, NY, 10603\\
$^b$Department of Statistical Science, Duke University, Durham, NC
27708-0251\\
$^c$\emph{Eunice Kennedy Shriver} National Institute of Child Health
\&
Human Development,\\
 National Institutes of Health, Bethesda, MD
20892 }

\newpage
\begin{abstract}
\noindent In many applications, it is of interest to assess the
dependence structure in multivariate longitudinal data. Discovering
such dependence is challenging due to the dimensionality involved.
By concatenating the random effects from component models for each
response, dependence within and across longitudinal responses can be
characterized through a large random effects covariance matrix.
Motivated by the common problems in estimating this matrix,
especially the off-diagonal elements, we propose a Bayesian approach
that relies on  shrinkage priors for parameters in a modified
Cholesky decomposition. Without adjustment, such priors and previous
related approaches are order-dependent and tend to shrink strongly
toward an AR-type structure. We propose moment-matching (MM) priors
to mitigate such problems. Efficient Gibbs samplers are developed
for posterior computation. The methods are illustrated through
simulated examples and are applied to a longitudinal epidemiologic
study of hormones and oxidative stress.

\vspace*{0.4cm} \noindent {\sc Key words}: Cholesky decomposition,
covariance matrix, moment-matching, oxidative stress, random
effects, shrinkage prior.

\end{abstract}

\newpage
\section{Introduction \label{section intro}}
In biomedical applications, there is increasing interest in the
analysis of multivariate longitudinal data, with \cite{Fieuws:2007}
providing a recent review of the literature in this area.  When the
dependence structure between different responses is not of interest,
one can potentially use marginal models for each response.
\cite{Gray:2000} use such an approach to combine inferences about a
treatment effect, using generalized estimating equations for model
fitting.  When the focus is instead on the time-varying relationship
between the different longitudinal responses, one can use a
multivariate random effects model, which allows correlations between
random effects in component models for each response \citep[][among
others]{Shah:1997,Chakraborty:2003}. \cite{Fieuws:2004} showed that
the random effects approach to joint modeling can sometimes produce
misleading results if the covariance structure is misspecified.

A well known problem that arises in fitting a joint random effects
model to multivariate longitudinal data is the presence of many
unknown parameters in the random effects covariance matrix.  This
makes standard methods for fitting random effects models subject to
convergence problems.  Even when the covariance matrix can be
estimated, the estimate tends to have a large variance and typical
methods do not allow for inferences on whether off-diagonal elements
of the random effects covariance are non-zero.  These issues lead to
difficulties in interpretation, which motivated \cite{Putter:2008}
to develop a latent class modeling approach.  In this article, we
instead attempt to improve the performance of the joint random
effects modeling approach through the use of a Bayesian method with
carefully-chosen priors placed on the covariance matrix to favor
sparsity.

This article is motivated by data from the BioCycle study, which
collected longitudinal measurements of markers of oxidative stress
and reproductive hormones over the menstrual cycle
\citep{Wactawski-Wende:2009}.  The goal is to improve understanding
of the dynamic relationship between these variables as this
relationship has complicated studies in women of reproductive age
with adverse health effects attributable to oxidative stress
\citep{Schisterman:2010}. In this study, fertility monitors were
used to time clinic visits and blood draws during two menstrual
cycles from 259 women \citep{Schisterman:2009}. Visits were
scheduled within each cycle during (1) menstruation, (2) mid
follicular phase, (3) late follicular phase, (4) luteinizing hormone
(LH) /follicle stimulating hormone (FSH) surge, (5) ovulation, (6)
early luteal phase, (7) mid luteal phase and (8) late luteal phase.
Serum samples were assayed for hormone levels including estradiol
(E2) and oxidative stress levels as measured by F2 Isoprostanes
(F2Iso). In this paper, we focus on investigating the relationship
between F2Iso, a biomarker of oxidative stress levels, and estradiol
(E2). The BioCycle Study provides a unique opportunity to study
dependence in hormone and oxidative stress trajectories.  Hormonal
patterns tend to follow patterns regulated by the
hypothalamic-pituitary-ovarian axis, and are strongly correlated
from cycle to cycle.

Following common practice for multivariate longitudinal data
analysis, we initially consider a linear mixed effects model
\citep{Laird:1982} for each response.  In particular, let $y_{hij}$
denote the measurement of response type $h$ for subject $i$ at visit
$j$, with $h=1$ for log-transformed E2 and $h=2$ for log-transformed
F2Iso, and $i=1,\ldots, n$, $j=1,\ldots, n_i$. Although our methods
focus on the bivariate case, they apply directly to general
multivariate longitudinal response data. We allow for unequal number
and spacing  of visits for the different women, assuming the visits
are missing at random (MAR) \citep{Rubin:1976}. This assumption is
deemed appropriate based on discussions with the study
investigators, as it is unlikely that the missing scheduled visits
were related to the F2Iso and E2 measurements on the day of the
missed visit. Letting ${\bf x}_{hij}$ and ${\bf z}_{hij}$ denote the
$p \times 1$ and $q \times 1$ vectors of predictors, we assume
\begin{eqnarray}
y_{hij}&=& {\bf x}_{hij}' \bet + {\bf z}_{hij}'{\bf b}_{hi}
+\epsilon_{hij}, \quad {\bf b}_{hi} \sim \mbox{N}_q( {\bf 0},
\bOmega ),\quad \epsilon_{hij}\sim
\mbox{N}(0,\sigma^2),\label{eq:lme}
\end{eqnarray}
where $\bet$ is a vector of unknown fixed effects parameters, ${\bf
b}_{hi}$ is a vector of random effects and is assumed independent of
the measurement error $\epsilon_{hij}$, $\bOmega$ is the $q \times
q$ random effects covariance matrix that reflects the dependence
structure within and across responses, and $\sigma^2$ is the
residual variance.

\begin{figure*}
\centering
\begin{tabular}{c}
    \includegraphics[width=4in]{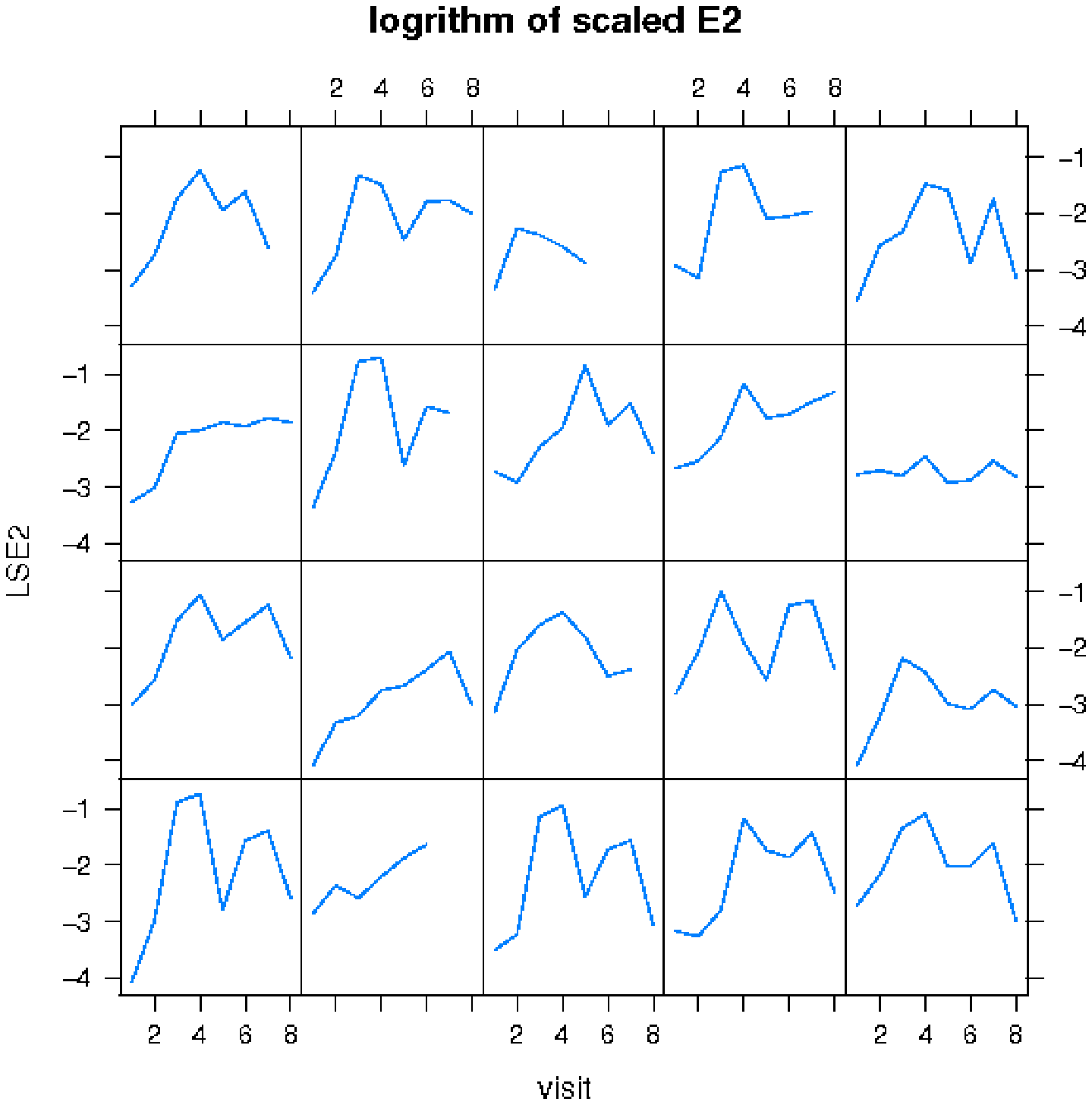} \\
   \includegraphics[width=4in]{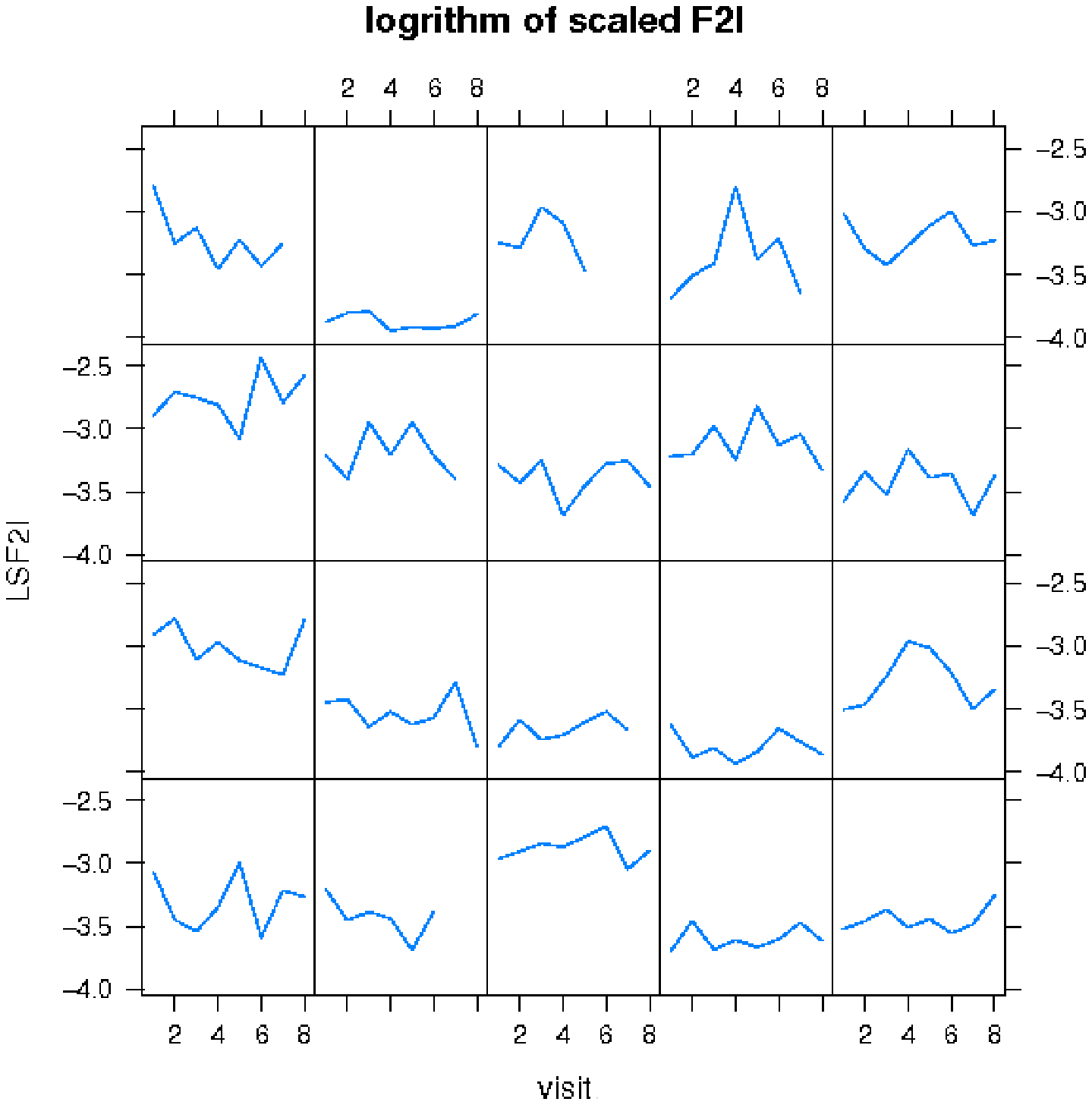}

\end{tabular}
\bigskip
 \caption{\small The first $20$ subjects in the data set. Responses LSE2 (logarithm of scaled E2) and LSF2I (logarithm of scaled F2Iso) are shown over visits 1-8. }
 \label{fig:20sub}
\end{figure*}

The joint mixed effects model (\ref{eq:lme}) is flexible in allowing
separate fixed and random effects for each response through the
appropriate choice of ${\bf x}_{hij}$ and ${\bf z}_{hij}$, while
accommodating dependence in the longitudinal trajectories through
dependence in the random effects.  Such dependence is measured by
the off-diagonal elements in the random effects covariance matrix
$\bOmega$. In the BioCycle study, there is substantial variability
in both F2Iso and E2  across the menstrual cycle as shown in Figure
\ref{fig:20sub}. Prior substantive knowledge suggests that the
trajectories of F2Iso and E2 over the cycle may differ for different
women, especially by menopausal status and body fat distribution.
Although we expect the patterns to be more similar among women in
the BioCycle study who were selected into the study because they
were healthy and regularly menstruating, there still exists
considerable variability. Hence, when studying certain populations
it may not be reasonable {\em a priori} to assume a simple
parametric model, such as a random intercept model. We instead
assume separate fixed and random effect coefficients for each visit.
This results in $p=9$ (intercept and coefficients for the eight
visits from each woman) and $q = 16$ (total number of responses if
the woman attended all of her scheduled visits for the two cycles),
for a total of $16\times15/2+9=129$ fixed and random effects
parameters to be estimated from the data of only 259 women.

In addition to the well-known problems of estimating a large number
of parameters without regularization, frequentist fitting of linear
mixed models with large numbers of random effects encounters
computational problems in requiring many inversions of a large
covariance matrix. The covariance matrix estimate is often
ill-conditioned in such cases, with the ratio between the largest
and smallest eigenvalues being large. This leads to amplification of
numerical errors when the matrix is inverted, resulting in either a
lack of convergence or apparent convergence to a poor estimate
having substantial bias and high variance. In fact, we first
attempted to fit this model using a standard frequentist approach
implemented in \textbf{R} 2.10.1 with the \emph{lme}() function
\citep{Pinheiro:1996,Lindstrom:1988}, but failed to obtain
convergence for the BioCycle data.

Given these problems, and our interest in inferences on certain
off-diagonal elements of the random effects covariance matrix
$\bOmega$, we instead adopt a Bayesian approach.   The typical
Bayesian approach to linear mixed effects models
\citep[e.g.,][]{Zeger:1991,Gilks:1993}, either assumes {\em a
priori} independence among the random effects or chooses an
inverse-Wishart prior distribution for the random effects covariance
structure. However, since the inverse-Wishart prior incorporates
only a single degree of freedom, it is not flexible enough as a
shrinkage prior for a high-dimensional covariance matrix. One
natural solution is to choose a prior that favors sparsity,
shrinking most insignificant elements of the covariance matrix to
values close to zero. This can stabilize estimation and improve
inferences on significant dynamic correlations.

A variety of the shrinkage priors for $\bOmega$ have been proposed
in the literature, achieving model flexibility while not sacrificing
the positive definite constraint through the use of matrix
decompositions. \cite{Daniels:1999} proposed priors that favor
shrinkage towards a diagonal structure.  \cite{Daniels:2002}
developed alternative priors based on a Cholesky decomposition,
giving advantages in interpretation and computation.
\cite{Smith:2002} proposed a parsimonious covariance estimation
approach for longitudinal data that avoids explicit specification of
random effects. Motivated by the problem of selecting random effects
with zero variance, \cite{Chen:2003} proposed a modified Cholesky
decomposition that facilitates choice of conditionally-conjugate
priors. \cite{Pourahmadi:2007} demonstrated appealing properties of
the \cite{Chen:2003} decomposition in terms of separation of the
variance and correlation parameters.

However, we find that posterior computation for the previously
proposed sparse shrinkage priors generally does not scale well as
the number of random effects increases and there are issues in
overly-favoring shrinkage towards AR-type covariance structures.
Motivated by the multivariate longitudinal BioCycle data, we propose
a new class of heavy-tailed shrinkage priors on the parameters in
the \cite{Chen:2003} decomposition. These priors are robust and
introduce substantial computational advantages. It is noted that
shrinkage priors under Cholesky-type decomposition have
computational advantages but induce order dependence and  tend to
over-shrink as the locations of the covariance matrix move further
off the diagonal. To mitigate this problem, we propose
moment-matching priors. Efficient Gibbs samplers are developed for
posterior inferences under both priors.

In Section 2, we describe the modified Cholesky decomposition of the
covariance matrix and propose new shrinkage priors for the
parameters in this decomposition. In Section 3, we describe the
order dependence phenomenon and propose the moment-matching priors.
Section 4 outlines a simple Gibbs sampling algorithm for posterior
computation. Section 5 applies the methods to simulated datasets.
 Section 6 considers the application to the BioCycle study and Section 7 concludes with a discussion.

\section{Shrinkage Priors for Random Effects Covariance Matrices}
In order to carry out a Bayesian analysis of model (\ref{eq:lme}),
we adopt the modified Cholesky decomposition of the covariance
matrix $\bOmega$ by \cite{Chen:2003},
\begin{eqnarray}
\bOmega = \bLambda \bGamma \bGamma' \bLambda,\label{eq:decomp}
\end{eqnarray}
where $\bLambda = \mbox{diag}( \lambda_1,\ldots, \lambda_q)$ is a
diagonal matrix with $\lambda_l \geq 0$ for $l=1,\ldots, q$, and
$\bGamma$ is a $q \times q$ unit lower triangular matrix with
$\gamma_{ml}$ in entry ($m, l$). The diagonal elements of $\bLambda$
and the lower triangular elements of $\bGamma$ are vectorized as
follows,
\begin{eqnarray*}
\blambda=(\lambda_1, \ldots, \lambda_q)', \quad
\bgamma=(\gamma_{21},\gamma_{31},\gamma_{32},\ldots, \gamma_{q,q-2},
\gamma_{q,q-1})'.
\end{eqnarray*}
The elements of $\blambda$ are proportional to the standard
deviations of the random effects. Setting $\lambda_l \approx 0$ is
effectively equivalent to excluding the $l$th random effect from the
model. By doing so, we move between models of different dimensions,
while keeping the covariance matrix of the random effects in each of
these models positive definite. The elements of $\bgamma$
characterize the correlations between the random effects.

Reparameterizing (\ref{eq:lme}) with the modified Cholesky
decomposition, we have
\begin{eqnarray}
y_{hij}&=& {\bf x}_{hij}' \bet + {\bf z}_{hij}'\bLambda\bGamma{\bf
a}_{hi} +\epsilon_{hij},\quad   {\bf a}_{hi} \sim \mbox{N}_q( {\bf
0}, {\bf I}_q ),\quad \epsilon_{hij}\sim
\mbox{N}(0,\sigma^2),\label{eq:lme2}
\end{eqnarray}
where ${\bf I}_q$ denotes a $q \times q$ identity matrix. Following
\cite{Chen:2003}, we define two vectors
\begin{eqnarray*}
u_{hij} = (a_{hil}\lambda_mz_{hijm})', \quad  t_{hij} =
\bigg\{z_{hijl}\big(a_{hil}+\sum_{m=1}^{l-1}a_{him}\gamma_{ml}\big)\bigg\}',
\quad 1\leq l<m\leq q.
\end{eqnarray*}
Then (\ref{eq:lme2}) can be rewritten as,
\begin{eqnarray}
y_{hij}-{\bf x}_{hij}'\bet-\sum_{l=1}^q a_{hil}\lambda_lz_{hijl}&=&
u_{hij}'\bgamma+\epsilon_{hij}, \label{eq:lme3}\\
y_{hij}-{\bf x}_{hij}'\bet&=&
t_{hij}'\blambda+\epsilon_{hij}.\label{eq:lme4}
\end{eqnarray}
Therefore prior distributions for $\bOmega$ can be induced through
priors on $\blambda, \bgamma$ and all the model parameters can be
updated as in the normal linear regression.

We first introduce the priors for the fixed effects (covariates)
coefficients $\bet$. When the number of covariates is large,
subset-selection is often desirable. In the Bayesian literature,
this is usually achieved by introducing a latent variable $J_l \in
\{0,1\}$ for each covariate that indicates whether it is included in
the model, and assuming a \emph{spike and slab} prior for $\beta_l$
conditional on $J_l$ \citep{George:1997,Smith:1996}. Let
$\bet^J=\{\beta_l: J_l=1\}$ be the set of coefficients of the
selected fixed effects and ${\bf X}^J$ be the corresponding
covariates matrix. We assume a standard i.i.d. Bernoulli prior for
$J_l$: $J_l\sim \mbox{Bernoulli}(p_0)$, and express prior ignorance
by setting $p_0\sim \mbox{beta}(a_p, b_p)$. Then, for each of the
$\beta_l$'s with $J_l=0$, we assume the prior to be a point mass at
$0$; and for $\bet^J$, we assume a Zellner g-prior
\citep{Zellner:1980},
\begin{eqnarray*}
\bet^J \sim  \mbox{N}(0,\sigma^2({\bf X}^{J^T}{\bf X}^J)^{-1}/g),
\quad g \sim \mbox{G}({1/2}, {N/2}), \label{eq:beta}
\end{eqnarray*}
where $\sigma^2$ follows a Jeffrey's prior $\sigma^2 \propto
1/\sigma^2$ is the same $\sigma^2$ in model (\ref{eq:lme}), $N=\sum
n_i$ and $\mbox{G}(a,b)$ denotes a Gamma distribution with mean
$a/b$ and variance $a/b^2$.

For  $\blambda$, we consider another point mass mixture prior
similar to that of the $\bet$'s, allowing for random effect
selection. Specifically, we assume an i.i.d. zero-inflated
half-normal distribution for $\lambda_l ~ (l=1,...,q)$,
\begin{eqnarray}
\lambda_l|\phi_l,p_l \sim p_l \delta_0+(1-p_l)\mbox{N}^+(0,
\sigma^2\phi_l^2), \quad \phi_l^2 \sim \mbox{IG}(1/2, 1/2),
\label{ZI}
\end{eqnarray}
where $\delta_0$ is a point mass at 0 and $\mbox{N}^+(0, \phi_l^2)$
is the normal distribution $\mbox{N}(0, \phi_l^2)$ truncated to its
positive support. When $\lambda_l>0$ for all $l$, the decomposition
in (\ref{eq:decomp}) guarantees that $\bOmega$ is positive definite
and $\bLambda$ and $\bGamma$ are identifiable. When $\lambda_l=0$,
elements of the resulting $\bOmega$ in the $l$th row and $l$th
column are 0. The submatrix of $\bOmega$ formed by removing the
$l$th row and $l$th column will still be positive definite.
Therefore we are able to move between models with different
dimensions by removing these rows and columns while still keeping
the covariance matrix of the random effects of all these models
positive definite. The hyperparameter $p_l$ represents the prior
probability of $\lambda_l=0$ and is set to be $0.5$ to express prior
ignorance. The induced marginal prior for $\lambda_l$ from
(\ref{ZI}) is a mixture of a  heavy-tailed truncated Cauchy
distribution and a point mass at zero.

The parameters of primary interest in this study are the
correlations of the random effects, which depend on $\bgamma$.
Without restriction, the large number of unknown parameters in
$\bgamma$ relative to the sample size can lead to difficulty in
model fitting. We thus consider the following
Normal-Exponential-Gamma (NEG)  shrinkage prior
\citep{Griffin:2007}:
\begin{eqnarray}
\gamma_{ml}|\psi_{ml} \sim \mbox{N}(0,\sigma^2\psi_{ml}),\quad
\psi_{ml}\sim \mbox{Exp}({\delta^2/2}),\quad \delta^2 \sim \mbox{G}
(c_0,d_0).\label{NEG}
\end{eqnarray}
The hyperparameters $(c_0, d_0)$ control the degree of model
sparsity. A larger $c_0$ and/or a smaller $d_0$ lead more
coefficients to be close to zero. The prior has fatter tails and
larger variance as $d_0$ increases. We set $c_0=1$ to introduce more
shrinkage and let $d_0\sim \mbox{G}(1,1)$ to make the priors more
flexible.

\section{Moment Matching Prior}
As noted in \cite{Pourahmadi:2007}, a perceived order among the
variables is central to the statistical interpretations of the
entries of $\bLambda$ and $\bGamma$ as certain  prediction variances
and moving average coefficients. For longitudinal and functional
data there is a natural time-order, while for others, the context
may not suggest a natural order.  The intrinsic order dependence in
shrinkage priors based on Choleskey-type decompositions, including
not only \cite{Chen:2003} but also \cite{Daniels:2002}, favors
shrinkage towards an autoregressive-type covariance structure. Such
methods can over shrink non-zero covariance not close to the
diagonal.  This motivated us to develop the following MM prior to
mitigate such order dependence problems.

Let $\gamma_{[ml]}$ denote the $m$th and $l$th row of the lower
triangular matrix $\bGamma$ and $\mu_{[ml]}$ denote the
corresponding prior mean for $\gamma_{[ml]}$,
\begin{eqnarray*}
\gamma_{[ml]}=(\gamma_{m1},\ldots,\gamma_{m,m-1},
\gamma_{l1},\ldots,\gamma_{l,l-1}),
\quad \mu_{[ml]}&=&(\mu_{m1},\ldots,\mu_{m,m-1}, \mu_{l1},\ldots,\mu_{l,l-1}),\\
&& 1\leq l<m \leq q.
\end{eqnarray*}
Also denote the correlation matrix corresponding to $\bOmega$ by
$\brho$. \cite{Chen:2003} showed that $\rho_{ml}$, the $(m,l)$th
entry of $\brho$, is determined solely by $\gamma_{[ml]}$ as
follows,
\begin{eqnarray}
\rho_{ml}=h(\gamma_{[ml]})=\frac{\gamma_{ml}+\sum_{r=1}^{l-1}\gamma_{lr}\gamma_{mr}}{\sqrt{(1+\sum_{r=1}^{l-1}\gamma_{lr}^2)(1+\sum_{r=1}^{m-1}\gamma_{mr}^2)}}.
\label{eq:rhogamma}
\end{eqnarray}
This property is crucial to the introduction of the MM prior. Our
key idea is to pair-wisely match the first and second prior moments
of $\brho$ to those induced from the priors for $\bgamma$'s. The
first order Taylor expansion of $h(\gamma_{[ml]})$ at the prior mean
of $\gamma_{[ml]}$ gives,
\begin{eqnarray}
h(\gamma_{[ml]})&\approx& h(\mu_{[ml]})+\nabla
h(\mu_{[ml]})'(\gamma_{[ml]}-\mu_{[ml]}), \label{eq:taylor1}
\end{eqnarray}
where $\nabla h(\mu_{[ml]})=(\frac{\partial h(\mu_{[ml]})}{\partial
\gamma_{m1}}, \ldots,\frac{\partial h(\mu_{[ml]})}{\partial
\gamma_{m,m-1}}, \frac{\partial h(\mu_{[ml]})}{\partial
\gamma_{l1}},\ldots,\frac{\partial h(\mu_{[ml]})}{\partial
\gamma_{l,l-1}})'$. Applying the expectation operator with respect
to the prior distribution of $\gamma_{[ml]}$ to (\ref{eq:taylor1}),
we have
\begin{equation}
\mbox{E}(\rho_{ml})=\mbox{E}\{h(\gamma_{[ml]})\}\approx
h(\mu_{[ml]})=\frac{\mu_{ml}+\sum_{r=1}^{l-1}\mu_{lr}\mu_{mr}}{\sqrt{(1+\sum_{r=1}^{l-1}\mu_{lr}^2)(1+\sum_{r=1}^{m-1}\mu_{mr}^2)}},
\quad 1\leq l<m\leq q. \label{eq:meanmatch}
\end{equation}
Fixing the values of $\mbox{E}(\rho_{ml})$'s and replacing the
approximation by equation (\ref{eq:meanmatch}), we define a system
of $q(q-1)/2$ equations for the prior means $\mu_{[ml]}$'s.
Similarly, applying the variance operator to (\ref{eq:taylor1}), we
have
\begin{equation*}
\mbox{Var}(\rho_{ml})=\mbox{Var}\big\{h(\gamma_{[ml]})\big\}\approx
\nabla h( \mu_{[ml]})' \Psi_{\gamma_{[ml]}} \nabla h( \mu_{[ml]}),
\quad 1\leq l<m\leq q,
\end{equation*}
where $\Psi_{\gamma_{[ml]}}$ is the prior covariance matrix of
$\gamma_{[ml]}$, with the variance of $\gamma_{mk}$ denoted by
$\psi_{mk}$ and the covariance between $\gamma_{lj}$ and
$\gamma_{mk}$ denoted by $\psi_{lj,mk}$. Rewriting the matrix
product in the form of summations and replacing the approximation by
the equation above, we have
\begin{equation}
\mbox{Var}(\rho_{ml})= \left \{
\begin{array}{cll}
\sum_{k=1}^{m-1}\psi_{mk}\big(\frac{\partial h(\mu_{[ml]})}{\partial
\mu_{mk}}\big)^2 +2\sum_{1\leq k<j}^{m-1}\psi_{mk,mj}\frac{\partial
h(\mu_{[ml]})}{\partial \mu_{mk}} \frac{\partial
h(\mu_{[ml]})}{\partial \mu_{mj}},& \mbox{for} \quad l=1, \\
\mbox{Var}(\rho_{l1})+\mbox{Var}(\rho_{m1})
+2\sum_{j=1}^{l-1}\sum_{k=1}^{m-1}\psi_{lj,mk} \frac{\partial
h(\mu_{[lj]})}{\partial \mu_{lj}} \frac{\partial
h(\mu_{[mk]})}{\partial \mu_{mk}}, & \mbox{for} \quad 2\leq l<m\leq
q,
\end{array}
\right. \label{eq:varmatch}
\end{equation}
where
\begin{equation*}
\frac{\partial h(\mu_{[ml]})}{\partial \mu_{mk}}= \left\{
\begin{array}{cl}
d_{ml}\big[\mu_{lk}
(1+\sum_{r=1}^{m-1}\mu_{mr}^2)-(\mu_{ml}+\sum_{r=1}^{l-1}\mu_{lr}\mu_{mr})\mu_{mk}\big],
&  \mbox{for} \quad k <l<m, \\
d_{mk}\big[(1+\sum_{r=1}^{m-1}\mu_{mr}^2)-\mu_{mk}(\mu_{mk}+\sum_{r=1}^{k-1}\mu_{kr}\mu_{mr})\big],&
\mbox{for} \quad k=l<m,\\
-d_{ml}\mu_{mk}(\mu_{ml}+\sum_{r=1}^{l-1}\mu_{lr}\mu_{mr}),&
\mbox{for} \quad l < k < m,
\end{array}
\right.
\end{equation*}
with
$d_{ml}=(1+\sum_{r=1}^{l-1}\mu_{lr}^2)^{-1/2}(1+\sum_{r=1}^{m-1}\mu_{mr}^2)^{-3/2}$.
When $\mbox{Var}(\rho_{ml})$'s and $\mbox{E}(\rho_{ml})$'s are
pre-fixed, (\ref{eq:varmatch}) defines a system of $q(q-1)/2$
equations for the prior covariances $\Psi_{\gamma_{[ml]}}$'s.

Lacking prior  information on the random effects, it is reasonable
to assume that all elements of the correlation matrix $\brho$ have
equal mean and variance \emph{a priori}, leaving the data to adjust
for the real correlations. If we assume a common prior mean $u$ and
variance $v$ for $\brho$'s, then $(u,v)$ should be in the range
$u\in [-1,1]$ and $u \pm 3\sqrt{v} \in [-1, 1]$ to satisfy the
condition $\rho_{ml}\in [-1,1]$. Solving (\ref{eq:meanmatch}), we
have,
\begin{equation}
\mu_{m1}= u\sqrt {\frac{1+(m-2)u}{\big(1+(m-1)u\big)(1-u)}}, \quad
\mu_{ml}=\mu_{m1}\sqrt{\frac{1-u}{\big(1+(l-2)u\big)\big(1+(l-1)u\big)}}.
\label{eq:mean}
\end{equation}
The system of equations (\ref{eq:varmatch}), however, is in general
under-identified because the number of unknowns is larger than the
number of equations. Under reasonable simplifying assumptions
motivated by the form of (\ref{eq:rhogamma}) and interpretations of
\cite{Pourahmadi:2007}, we assume that $\gamma_{m1}, \ldots,
\gamma_{m, m-1}$ are independent of each other, the $\gamma_{ml}'s$
have common variance, and the correlations between $\gamma_{lj}$ and
$\gamma_{mk}$ $(l \neq m)$ are equal.  Thus, the number of unknowns
and equations become the same and  unique solutions for $\psi_{m1}$
and $\psi_{m2}$ can be written with the above assumptions for all
$m$ by,
\begin{eqnarray}
\psi_{m1}&=&v/\sum_{k=1}^{m-1}\big(\frac{\partial
h(\mu_{[ml]})}{\partial \mu_{mk}}\big)^2 \nonumber\\
\psi_{m2}&=&-v/(2 \sum_{j=1}^{l-1}\sum_{k=1}^{m-1} \frac{\partial
h(\mu_{[lj]})}{\partial \mu_{lj}} \frac{\partial
h(\mu_{[mk]})}{\partial \mu_{mk}}).\label{eq:var}
\end{eqnarray}
Given the means and variances of $\brho$, we can calculate the
corresponding means and variances of $\gamma_{ml}$'s through
(\ref{eq:mean}) and (\ref{eq:var}). To test the effectiveness of the
transformation, we can generate $\gamma_{ml}'s$ 1000 times and
obtain the corresponding estimated prior distributions of $\brho$
 through (\ref{eq:rhogamma}). To set values for $(u,v)$, we
want to both shrink nonsignificant values as much as possible by
setting $u$ close to zero and leave out significant values by
setting $v$ as large as possible but within the constraint that
$u\in [-1,1]$ and $u \pm 3\sqrt{v} \in [-1, 1]$. With the above two
criteria, to test the effectiveness of the MM priors, we experiment
with different values of $(u, v)$, $(0.05, 0.1)$, $(0.1, 0.09)$,
$(0.15, 0.08)$, $(0.2, 0.07)$ with different dimensions. In all the
experiments, order-dependence is clearly avoided as the entries of
$\brho$ move further off the diagonal. The prior distributions are
still approximately $\mbox{N}(u,v)$. We notice that with $u=0.05$ or
0.1, the resulting elements of the estimated $\brho$ have relatively
larger ranges, while as $u$ increases, the range decreases. To
achieve more flexibility, we can set weakly-informative priors for
$u$ and $v$ as $u\sim \mbox{N}(\mu_0,\sigma_0^2)1(\mu \in [-1,1])$
and $v\sim \mbox{IG}(c_0,d_0)1(u\pm 3\sqrt{v}\in [-1,1])$
respectively. The corresponding priors for $\bmu$ and $\Psi$ can
then be calculated from model (\ref{eq:mean}) and (\ref{eq:var}) and
some Jacobian computation is needed.

\section{Posterior Inferences}
The posterior distribution is obtained by combining priors and the
likelihood in the usual way. However, direct evaluation of the
posterior distribution seems to be difficult.  The joint posterior
distribution for $\theta= (\bet, \blambda, \bgamma, \sigma^2)$ in
model (3) is given by,
\begin{eqnarray}
p(\theta|y)\propto \bigg[\prod_{i=1}^n  \mbox{N}_q ({\bf{a}}_i;0,
\mbox{I}_q)\prod_{h} \big\{\prod_{j=1}^{n_i}\mbox{N}(y_{hij};
{\bf{x}}_{hij}'\bet+{\bf{z}}_{hij}'\bLambda\bGamma {\bf{a}}_i,
\sigma^2)\big\}\bigg]p(\sigma^2)p(\bet, J, g) p(\blambda, \bgamma),
\label{eq:full}
\end{eqnarray}
which has a complex form that makes direct sampling infeasible.
Therefore we employ the Gibbs sampler \citep{Gelfand:1990} by
iteratively sampling from the full conditional distributions of each
parameter given the other parameters. The details of our Gibbs
sampler is given below:
\begin{enumerate}
\item Sampling fixed effects parameter $\bet$ through,
\begin{eqnarray*}
p(\bet^J|\cdots) &\sim& \mbox{N} (\mu_{\bet^J},\Sigma_{\bet^J}),
\end{eqnarray*}
where
$\Sigma_{\bet^J}=\frac{1}{g+1}(\sum_{h,i,j}\frac{1}{\sigma^2}x_{hij}^Jx_{hij}^{J'})^{-1}$
and  $ \mu_{\bet^J}= \Sigma_{\bet^J}
\sum_{h,i,j}\frac{1}{\sigma^2}x_{hij}^J \phi_{hij}$, with
$\phi_{hij}= y_{hij}-z_{hij}'\bLambda\bGamma {\bf a}_i$ and
$x_{hij}^J$ denoting the subvector of $x_{hij}$, $\{x_{hijl}:
J_l=1\}$.

\item  Sampling $g$ through the following conjugate Gamma
distribution,
\begin{eqnarray*}
g &\sim& \mbox{G}\bigg(\frac{p_J+1}{2}, \frac{\bet^{J'}
(\sum_{h,i,j}\frac{1}{\sigma^2}x_{hij}^{J'}x_{hij}^J)\bet^{J}+N}{2}\bigg),
\end{eqnarray*}
where $p_J=\sum_{l=1}^p1(J_l=1)$ and $N=\sum_{i}n_i$.

\item Updating $J_l$ individually,
following results from Smith and Kohn (1996), we have
\begin{eqnarray*}
p(J_l=1|\cdots)&=&\frac{1}{1+h_l}
\end{eqnarray*}
with $h_l=
\frac{1-p_l}{p_l}(1+\frac{1}{g})^{\frac{1}{2}}\{\frac{S(J_l=1)}{S(J_l=0)}\}^{\frac{N}{2}}$
and $S(J)= \phi'\phi-\frac{1}{1+g}\phi'{\bf X} ^J({\bf X}^{J'}{\bf
X}^J)^{-1}{\bf X}^{J'}\phi$.

\item Sampling $\lambda_l$ individually from a inflated
half-normal distribution with
$\zeta_{hij}=y_{hij}-x_{hij}^{J'}\bet^J$
\begin{eqnarray*}
p(\lambda_l|\cdots) &=&
\mbox{ZI-N}^{+}(\hat{p_l},\hat{\lambda_l},\hat{\sigma_l^2})
\end{eqnarray*}
with $\hat{p_l}=
\frac{p_l}{p_l+(1-p_l)\frac{N(0;0,1)}{N(0;\hat{\lambda_l},\hat{\sigma_l^2})}\frac{1-\Phi(0;\hat{\lambda_l},\hat{\sigma_l^2})}{1-\Phi(0;0,1)}}$,
$\hat{\lambda_l}=
\hat{\sigma_l^2}(\sum_{h,i,j}\frac{1}{\sigma^2}t_{hijl}(\zeta_{hij}-\sum_{k\neq
l}t_{hijk}\lambda_k)$ and
$\hat{\sigma_l^2}=(\sum_{h,i,j}\frac{t_{hijl}^2}{\sigma^2}+1)^{-1}$.

\item Updating $\bgamma$ through the following two circumstances,
\begin{enumerate}[i.]
\item If the prior is as described in (\ref{NEG}), following \cite{Park:2008}, we can use blocked Gibbs sampler to
update $\bgamma$'s and their concentration parameters as following,
\begin{eqnarray*}
\bgamma\sim \mbox{N}(\mu_{\bgamma},\Sigma_{\bgamma}), \quad
\frac{1}{\psi_l^2}\sim \mbox{Inverse-Gaussian}
(\sqrt{\frac{\delta^2}{\gamma_{ml}^2}},\delta^2), \\
\delta^2\sim
\mbox{G}(c_0+r,1/(\sum_{m,l}\frac{\gamma_{ml}^2}{2}+d_0)), \quad
d_0\sim \mbox{G}(1,1+\delta^2)
\end{eqnarray*}
with $\mu_{\bgamma}=
\Sigma_{\bgamma}\sum_{h,i,j}\frac{1}{\sigma^2}u_{hij}w_{hij}$ and
$\Sigma_{\bgamma}=
(\sum_{h,i,j}\frac{u_{hij}u_{hij}'}{\sigma^2}+D_{\psi}^{-1})^{-1}$.
$D_{\psi}$ is a diagonal matrix with diagonal elements of
$\psi_{ml}$.

\item If the prior is the MM prior, $\bgamma$ is
updated by,
\begin{eqnarray*}
p(\bgamma|\cdots) \sim \mbox{N}(\hat{\bgamma}, \hat{R})
\end{eqnarray*}
where
$\hat{R}=(\sigma^{-2}\sum_{h,i,j}u_{hij}u_{hij}^T+\Psi^{-1})^{-1}$
and
$\hat{\bgamma}=\hat{R}\{\sigma^{-2}\sum_{h,i,j}u_{hij}(y_{hij}-x_{hij}^{J'}
\bet^J)+\Psi^{-1}\bmu\}$. $\bmu$ and $\Psi$ are  obtained from the
MM priors described in Section 3 and can be  updated through the
random walk Metropolis-Hastings method if hyperpriors $u$ and $v$
are not fixed.

\end{enumerate}

\item Sampling random effects ${\bf a}_{hi}$ from
\begin{eqnarray*}
p({\bf a}_{hi}|\cdots)&\sim& \mbox{N} (\mu_{{\bf
a}_{hi}},\Sigma_{{\bf a}_{hi}}),
\end{eqnarray*}
with $\mu_{{\bf a}_{hi}}= \Sigma_{{\bf
a}_{hi}}\sum_{j}\frac{1}{\sigma^2}\bGamma'\bLambda \zeta_{hij} $ and
$ \Sigma_{{\bf a}_{hi}}=(\sum_{j}\frac{1}{\sigma^2}\bGamma'\bLambda
z_{hij}z_{hij}'\bLambda\bGamma+ \mbox{I}_q)^{-1}$.

\item Sampling $\sigma^2$ with $\theta_{hij}= y_{hij}-x_{hij}^{J'}\bet^J-z_{hij}'\bLambda\bGamma
{\bf a}_{hi}$ by,
\begin{eqnarray*}
p(\sigma^2|\cdots)\sim \mbox{Inverse-Gamma}(N/2,
\sum_{h,i,j}\theta_{hij}^2/2)
\end{eqnarray*}
\end{enumerate}
After discarding the draws from the burn-in period, we can estimate
posterior summaries of the model parameters in the usual way from
the Gibbs sampler output.

\section{Simulations}
\begin{figure}
\centering
\begin{tabular}{cc}
    \includegraphics[width=3in]{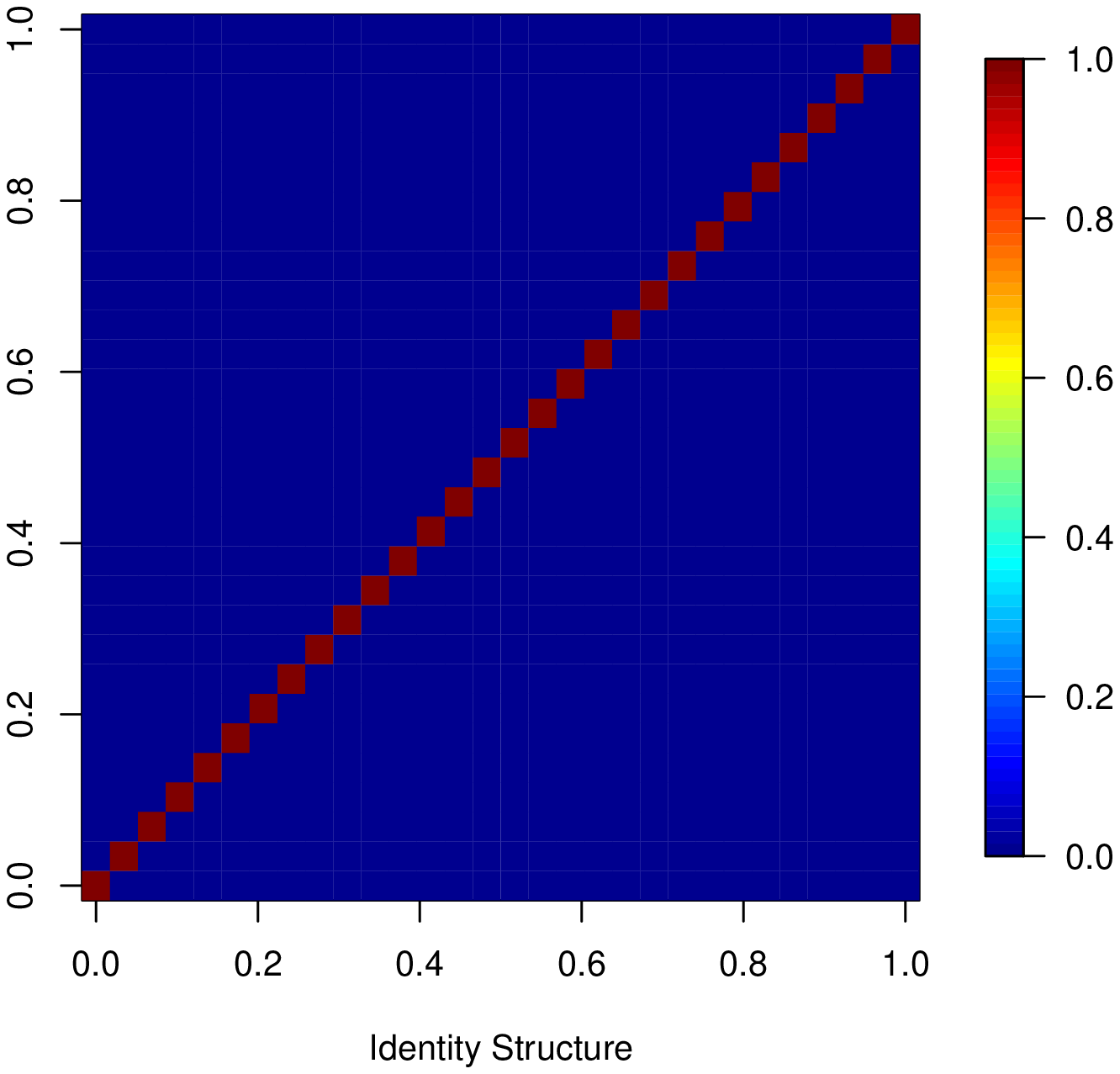}
    & \includegraphics[width=3in]{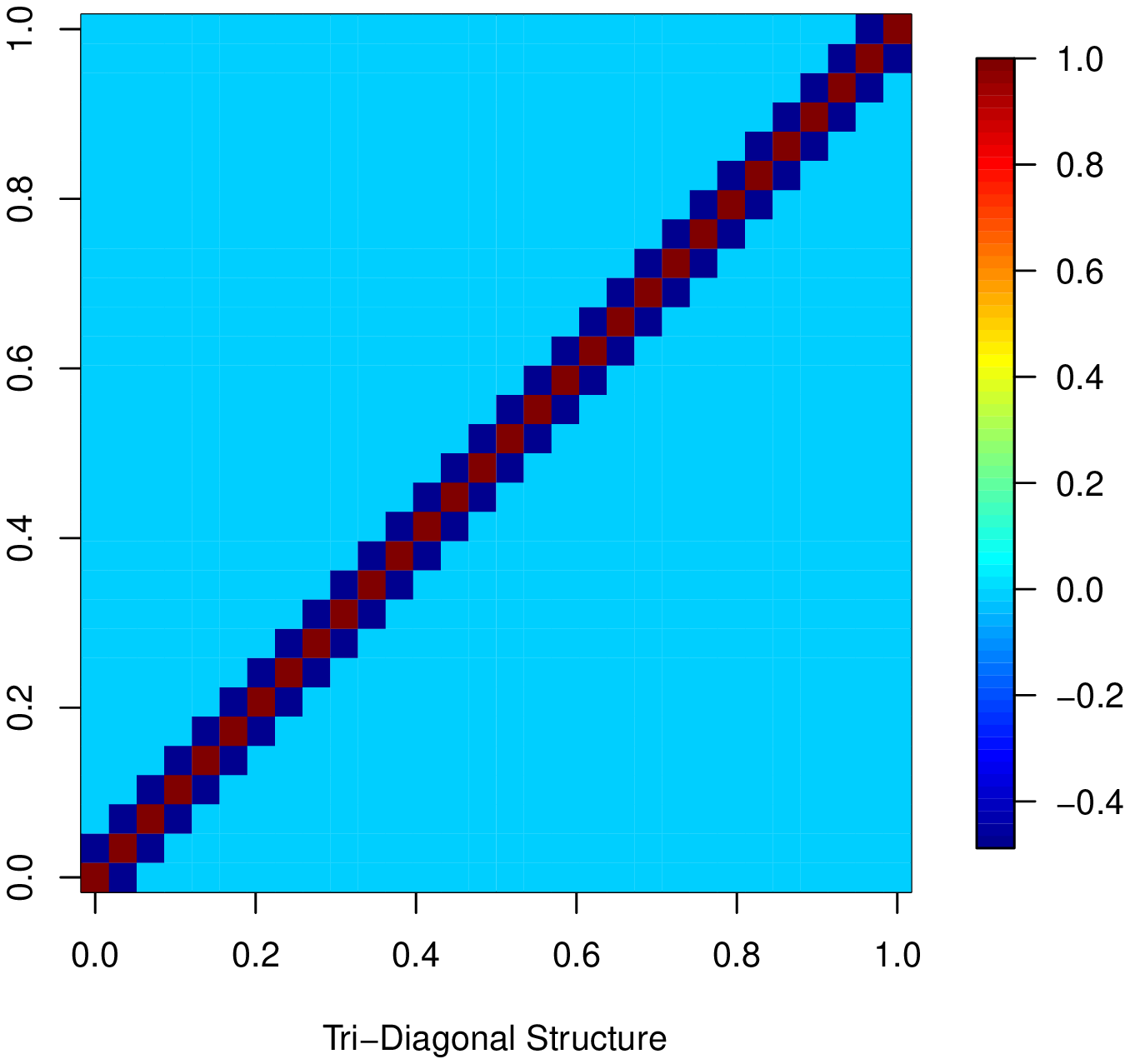} \\
    \includegraphics[width=3in]{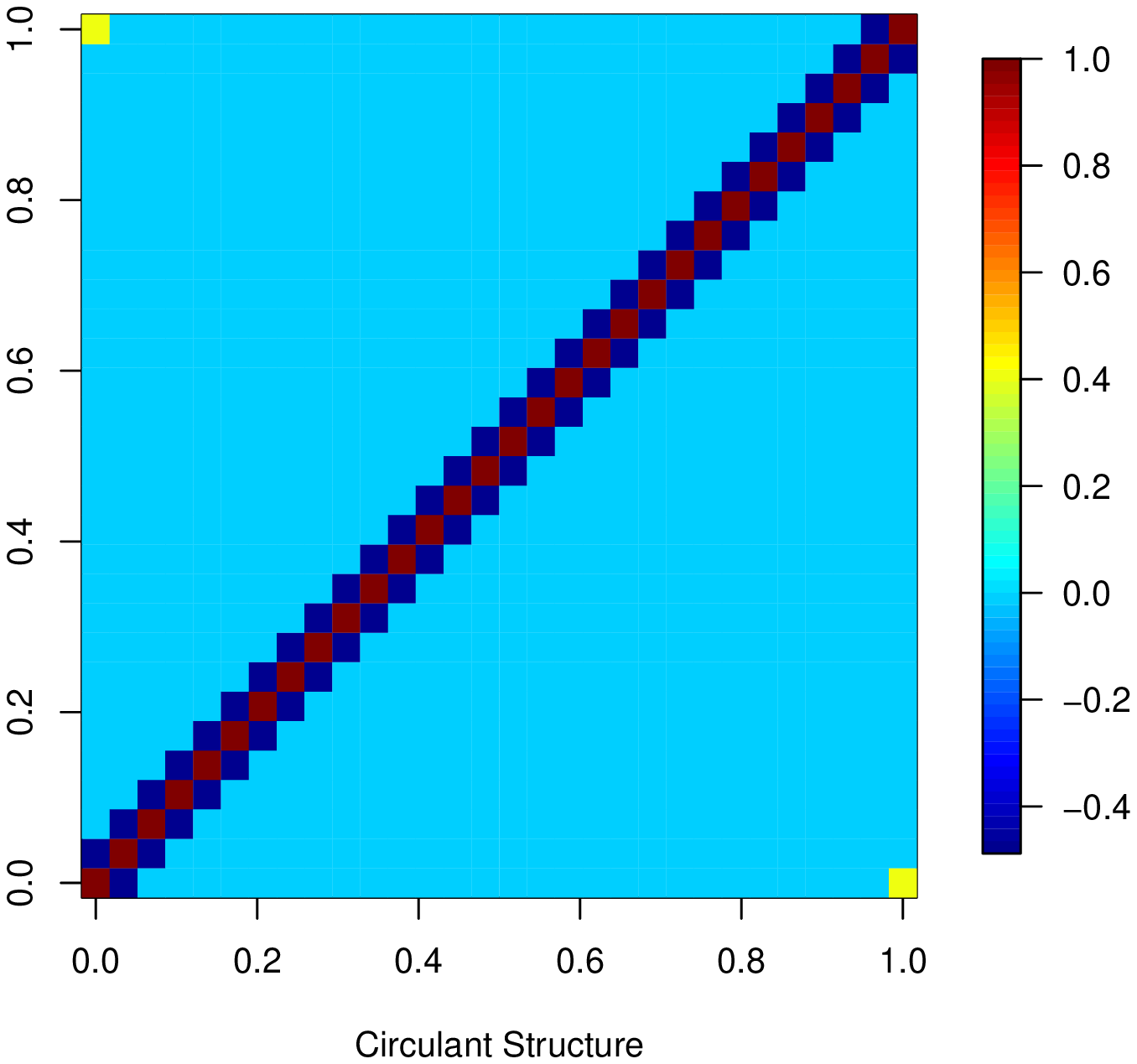} &   \includegraphics[width=3in]{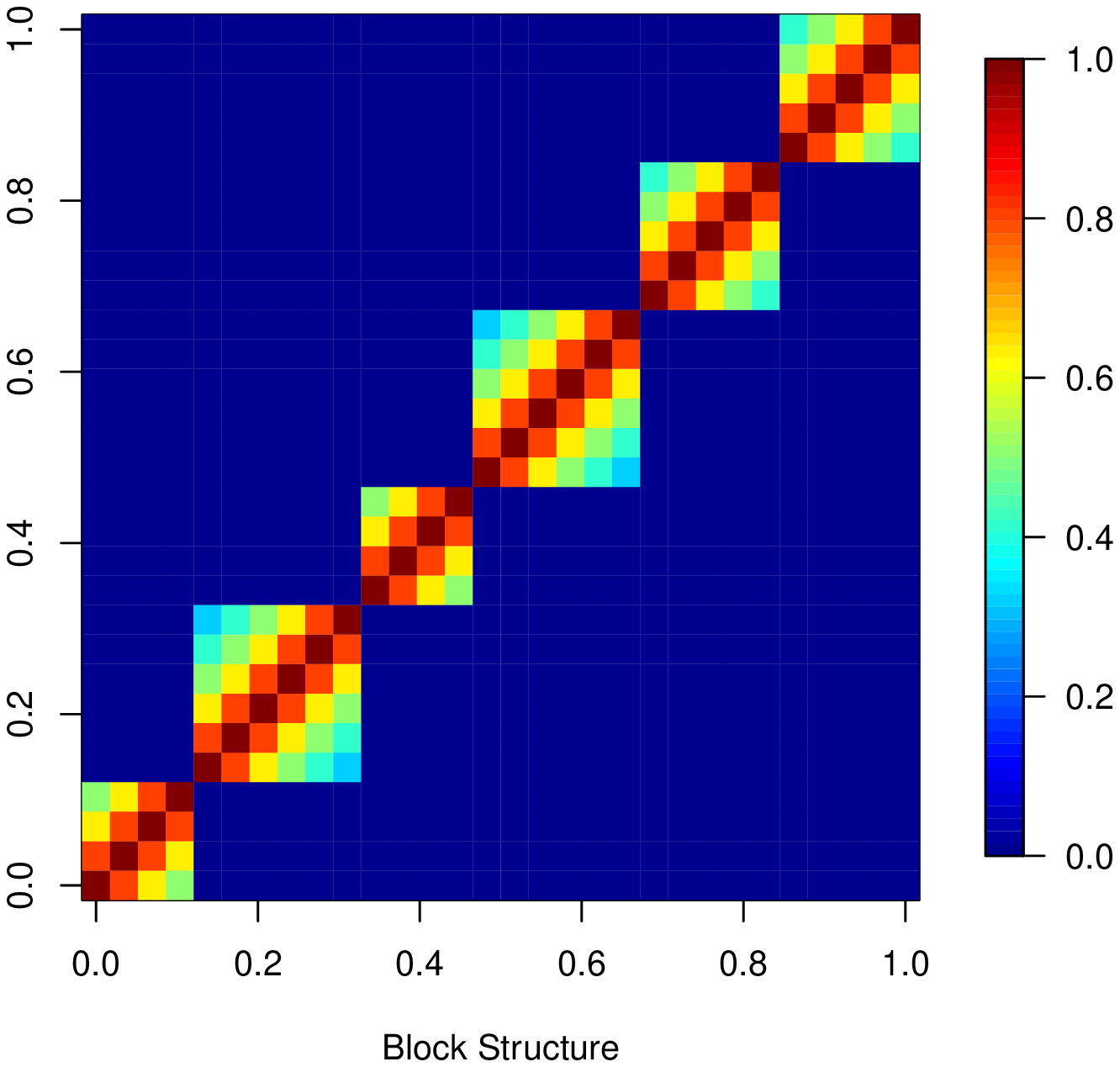}\\
    \includegraphics[width=3in]{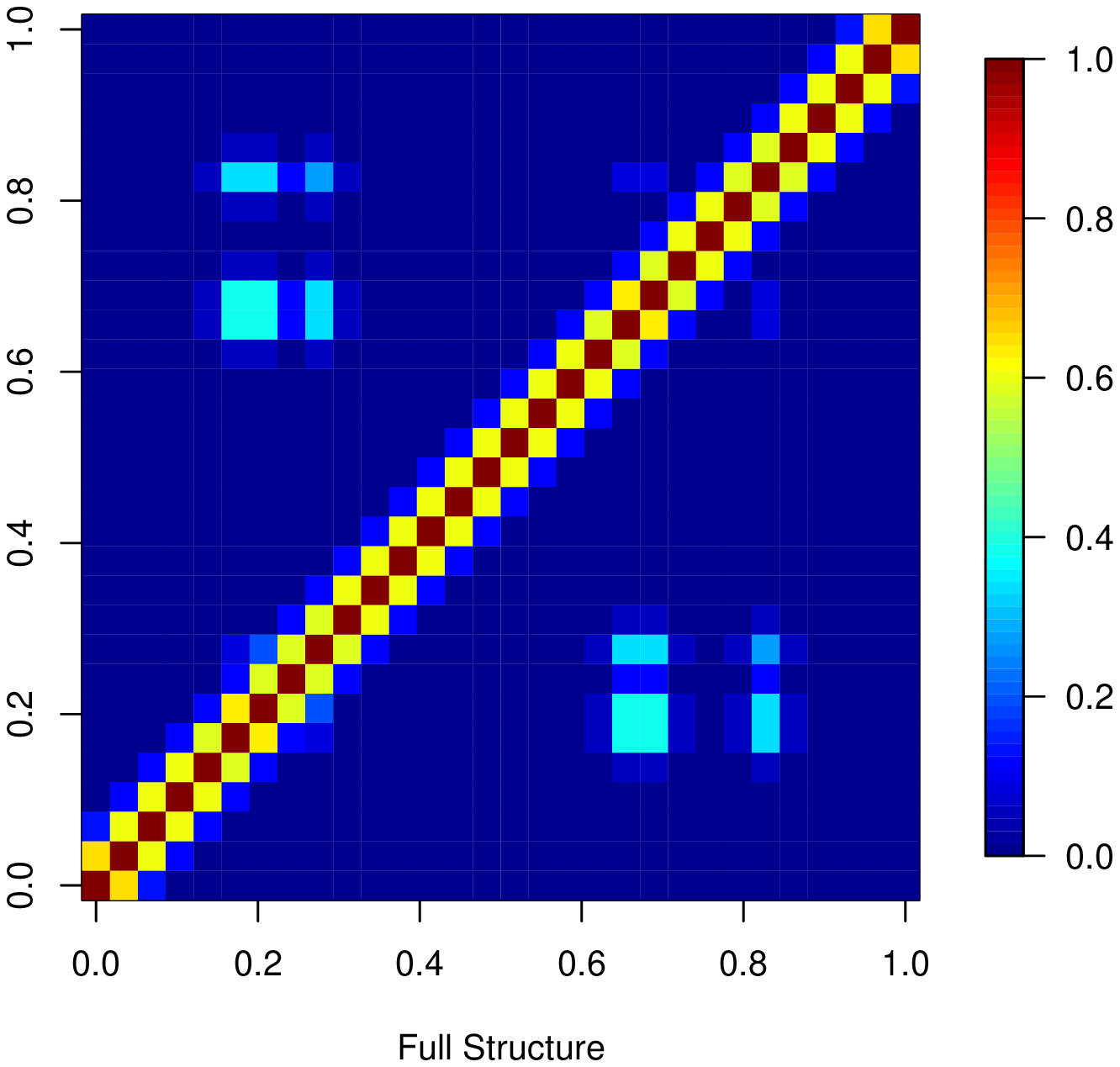} &   \includegraphics[width=3in]{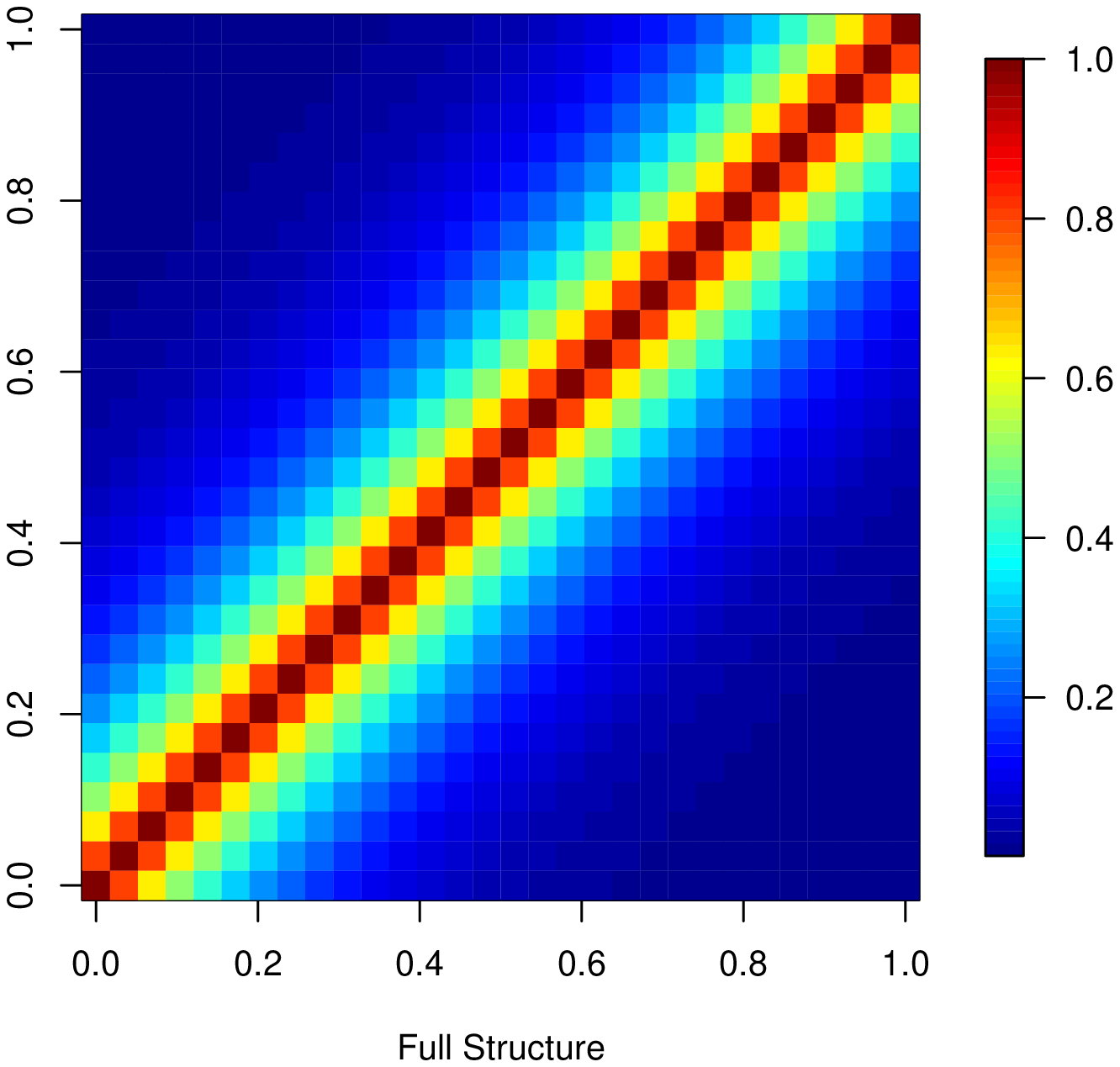}
\end{tabular}
\bigskip
 \caption{\small Imageplots for six true covariance matrices}\label{true}
\end{figure}

In this section, we examine the performance of the proposed priors
on simulated data. Since our primary interest is on the covariance
matrix of random effects, we assume there are no fixed effects in
the simulations. Data are generated from model (\ref{eq:lme2}) with
$z_{hij}=\mbox{I}_q$. Six representative structures (Figure
\ref{true}) are considered.

\begin{enumerate}
\item The \emph{identity} structure: $\bOmega$ is the identity
matrix so all random effects are independent.

\item The \emph{tri-diagonal} structure: $\bOmega$ has unity diagonal entries
with the immediate off-diagonal entries being -0.488, corresponding
to the covariance matrix of a MA(1) model with decay parameter 0.8.
The remaining entries are zero.

\item The \emph{circulant} structure: similar to the
tri-diagonal structure except for an additional pair of entries at
$(1,q)$ and $(q,1)$ being set to 0.4.

\item The \emph{block diagonal} structure: $\bOmega$ has six blocks, each viewed
as a separate covariance matrix with the entries decreasing from
unity at the rate of 0.8 as a function of the distance from the
diagonal (i.e., the immediate off-diagonal entries are 0.8 and the
next off- diagonals are $0.8^2$, and so on). This resembles the
situation where the variables are divided into several independent
groups and variables within the same group are closely connected.

\item The \emph{random} structure: $\bOmega$ has the diagonals being unity, the immediate off-diagonal entries being 0.4,
and some other entries having randomly selected values. We also
experiment with other values for the immediate off-diagonal entries.
This structure is similar to that in our application, where the data
are longitudinal but can have significant points further off the
diagonal entries.

\item The \emph{full} structure: similar to the block diagonal
structure, but all variables are now in the same group. Entries
decay at the rate of 0.8 as they swing away from the main diagonals,
resembling an AR(1) structure.
\end{enumerate}
For each structure, we simulate a data set with 200 subjects, each
having 15 visits and 2 outcomes per visit. In total, there are 30
random effects for each subject, i.e., $q=2\times15=30$.

\begin{figure}
\centering
\begin{tabular}{c}
    \includegraphics[width=3in]{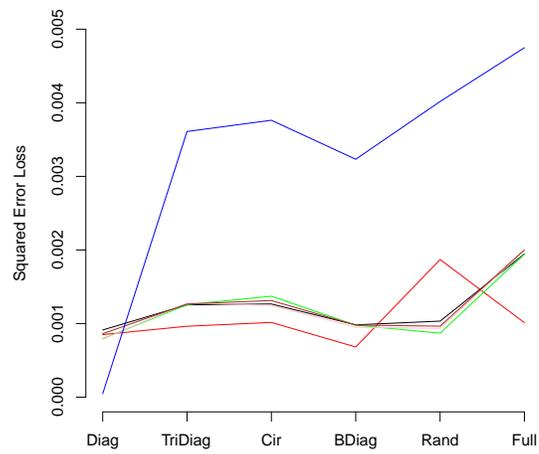}
\end{tabular}
\bigskip
 \caption{\small Squared Error Loss for Maximum Likelihood Estimation(blue line), Moment Matching Prior with $(u,v)=(0.05, 0.1)$(black line), $(u,v)=(0.1, 0.09)$(green line), $(u,v)=(0.15,0.08)$(pink line), $(u,v)=(0.2,0.07)$(brown
 line)  and Shrinkage prior(red line)}\label{compare}
\end{figure}

We first try to estimate the model with functions from the
\textbf{R} package \emph{nlme}, which can fit and compare Gaussian
linear and nonlinear mixed-effects models. We can only get
estimation when the covariance matrix is diagonal, while all the
others fail with an error message ``iteration limit reached without
convergence". It seems that the package \emph{nlme} can only deal
with small dimensional data, e.g., q is small. We then try the
\textbf{R} package \emph{corpcor} for comparison with our proposed
methods. This package implements a James-Stein-type shrinkage
estimator for the covariance matrix, with separate shrinkage for
variances and correlations. The details of the method are explained
in \cite{Schafer:2005} and \cite{Opgen-Rhein:2007}. In order to
compare the estimated covariance matrix with different methods (the
\textbf{R} package \emph{corpcor}  cannot output the covariance
matrix for the random effects in the linear mixed effects model), we
assume that the residual variance is zero when generating the data.
Results with the shrinkage and the MM priors are based on a Gibbs
sampler of 20,000 iterations after a burn-in period of 10,000.
Estimations are compared based on the squared error loss function,
\begin{eqnarray*}
D(\hat{\bOmega},
\bOmega)=\frac{1}{q^2}\big\{\sum_i\sum_j(\hat{\omega}_{ij}-\omega_{ij})^2\big\}^{1/2}.
\end{eqnarray*}
where $\omega_{ij}$ is in the $i$th row, $j$th column of $\bOmega$
and $\hat{\omega}$ is in the $i$th row, $j$th column of
$\hat{\bOmega}$. Figure \ref{compare} shows the squared error losses
of the estimates from the \emph{corpcor}, the shrinkage priors and
the MM priors. For simplicity, we set fixed values for $(u,v)$ as
$(0.05,0.1)$, $(0.1,0.09)$, $(0.15, 0.08)$ and $(0.2, 0.07)$. Both
the shrinkage priors and the MM priors outperform the  estimation
from the \emph{corpcor},  except under the diagonal covariance
matrix structure. The  \emph{corpcor}  performs best when the true
underlying covariance matrix is sparse but otherwise tends to
over-shrink.  When the true underlying covariance structure is
diagonal, the shrinkage priors and the MM priors perform equally
well. The shrinkage priors have the smallest squared error losses
when the underlying covariance structure is tri-diagonal, circulant,
block-diagonal and full structure. The MM priors clearly outperform
the shrinkage priors when the true underlying covariance structure
is random.  As expected, the estimates of the shrinkage priors under
the random structure tend to over-shrink the parameters as they move
further off the diagonal.

\begin{figure}
\centering
\begin{tabular}{c}
    \includegraphics[width=3in]{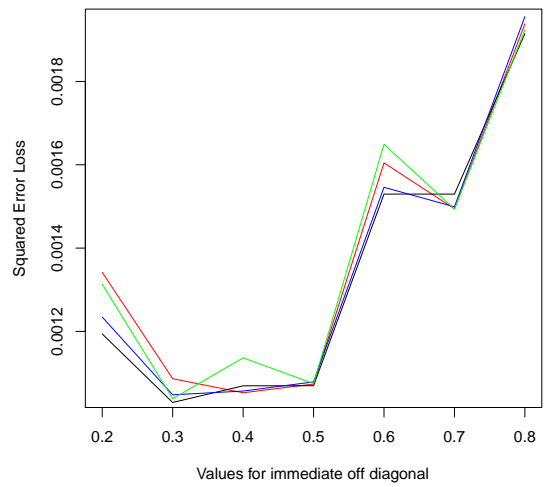}
\end{tabular}
\bigskip
 \caption{\small Squared Error Loss for $(u,v)=(0.05, 0.1)$(red line), $(u,v)=(0.1, 0.09)$(black line), $(u,v)=(0.15,0.08)$(blue line), $(u,v)=(0.2,0.07)$(green line)}\label{srl}
\end{figure}

To further explore the impact of $(u,v)$ values on the performance
of the MM priors, we calculate the MSE for $(u,v)$ values being
$(0.05,0.1), (0.1, 0.09), (0.15, 0.08), (0.2, 0.07)$ under different
random covariance structures. The difference in MSE from the MM
priors among the selected $(u,v)$ values under the above simulation
settings are very small. The immediate off-diagonal values are
chosen from $\{0.2, 0.3, \ldots, 0.8\}$ and the randomly selected
further-off diagonal values are the same in each test for
comparison. MSEs are shown in Figure \ref{srl} and the MM priors
perform best with $(u,v)=(0.1, 0.09)$.  Since $(u,v)$ are
hyperpriors for the correlation matrix, which will not be affected
by the magnitude and scale of the new datasets, we adopt the value
(0.1,0.09) in  later analyses for simplicity. MSEs are smallest when
the immediate off-diagonal value is 0.3 and get larger when the
values get larger.

\section{Application to the BioCycle Study}
Oxygen free radicals have been implicated in spontaneous abortions,
infertility in men and women, reduced birth weight, aging, and
chronic disease processes, such as cardiovascular disease and
cancer. It is thought that estrogen may play an important role in
oxidative stress levels in women. However, little is known about the
relation between oxidative stress, estrogen levels, and their
influence on outcomes, such as likelihood of conception or
spontaneous abortions. The primary goals of the BioCycle study are
to better understand the intricate relationship between hormone
levels and oxidative stress during the menstrual cycle
\citep{Schisterman:2010}. The BioCycle study enrolled 259 healthy,
regularly menstruating premenopausal women for two menstrual cycles.
Participants visited the clinic up to 8 times per cycle, at which
time blood and urine were collected.

The BioCycle study provides a unique setting for application of the
proposed methodology.  The data is longitudinal and hormone levels
tend to follow predicted patterns across the menstrual cycle due to
the complex feedback mechanisms which regulate hormonal levels
through the hypothalamic-pituitary-ovary axis.  Further, hormone
levels during specific phases tend to be correlated from cycle to
cycle.

The responses are transformed to a log scale to make the normal
assumption more reliable and the predictors are standardized by
subtracting the mean and dividing by the standard deviation.
Responses of F2Iso and E2 from the first 20 subjects are shown in
Figure \ref{fig:20sub}. We can see certain common trends over visits
across the women, but more strikingly each individual has her own
diversity which makes the plots more variable. Linear mixed-effects
models can accommodate such differences and analyze the longitudinal
dependences among two types of responses varying over visits through
the covariance matrix. Specifically, $y_{hij}$ is the response for
type $h$ ($h=1,2$) of subject $i$ ($i=1,\ldots,259$) at visit $j$
($j=1,\ldots,8$ for the 8 visits). Let
$x_{1ij}=(1,x_{i11},x_{i12},x_{i13},x_{i14},0,0,0,0)_{9\times1}^{'}$
and
$x_{2ij}=(1,0,0,0,0,x_{i21},x_{i22},x_{i23},x_{i24})_{9\times1}^{'}$
be the fixed predictors of response $h=1$ and $h=2$, respectively,
for subject $i$ at visit $j$. Let
$z_{1ij}=(\underbrace{0,\ldots,1_{1ij},\ldots,0}_{8\times1},\underbrace{0,\ldots,0}_{8\times1})_{16\times1}^{'}$
and $z_{2ij}=(\underbrace{0,\ldots,0}_{8\times1},
\underbrace{0,\ldots,1_{2ij},\ldots,0}_{8\times1})_{16\times1}^{'}$
stand for the random predictors of response $h$ for subject $i$ at
visit $j$, where
\begin{displaymath}
1_{hij}=\left\{ \begin{array} {lc}   1,& \mbox{subject $i$ showed up at visit $j$ for response $h$}, \\
                                \,   0,& \mbox{otherwise}.

                                \end{array}\right.
\end{displaymath}
We attempt to fit model (\ref{eq:lme}) with the \emph{lme}()
function in \textbf{R} 2.10.1 but failed, because the estimates do
not converge. We first estimate the covariance structure with the
shrinkage priors given the data collected longitudinally. In order
to capture the possible sporadic significant signals, we also
estimate model (\ref{eq:lme}) with the MM priors with $(u,v)=(0.1,
0.09)$. The Raftery and Lewis diagnostic \citep{Raftery:1995} is
used to estimate the number of MCMC samples needed for a small Monte
Carlo error in estimating the 95\% credible intervals. The required
sample size can be different for each parameter and 20,000
iterations are found to be enough for all parameters. Convergence
diagnostics, such as trace plots and Geweke's convergence diagnostic
for randomly selected off-diagonal elements of the covariance matrix
are performed on some selected elements. No signs of adverse mixing
is found. All results are based on 50,000 Gibbs sampling iterations
after a burn-in period of 20,000.

\begin{figure}
\centering
\begin{tabular}{cc}
    \includegraphics[angle=-90,width=3in, scale=1]{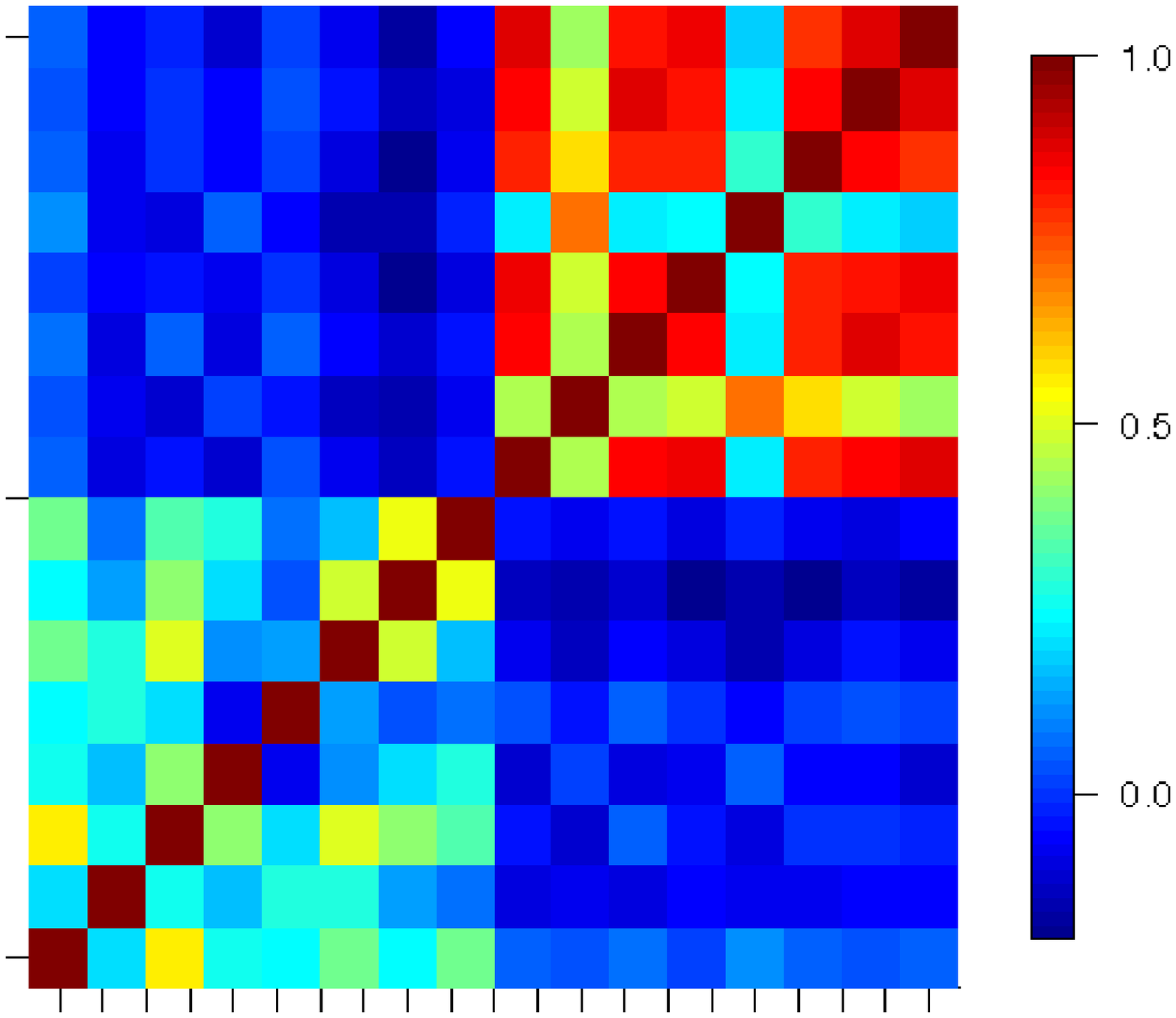} &
    \includegraphics[angle=-90,width=3in, scale=1]{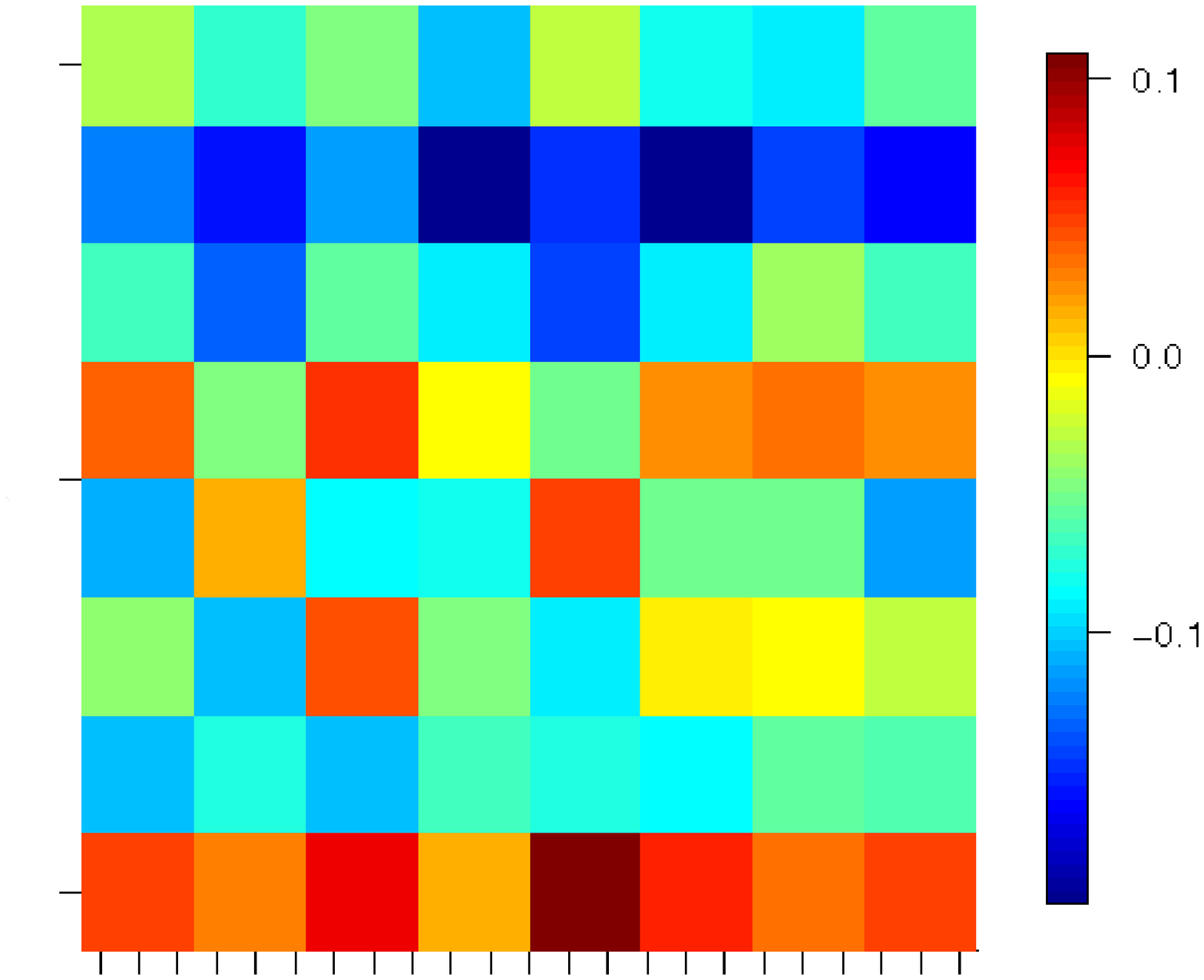}
\end{tabular}
\begin{minipage}{0.8\textwidth}
\bigskip
 \caption{\small Estimated covariance matrix between log(E2) and log(F2Iso) and the zoomed in cross covariance structure through Moment Matching prior }
 \label{fig:E2F2Iso1}
\end{minipage}
\end{figure}

\begin{figure}
\centering
\begin{tabular}{cc}
    \includegraphics[angle=-90,width=3in, scale=1]{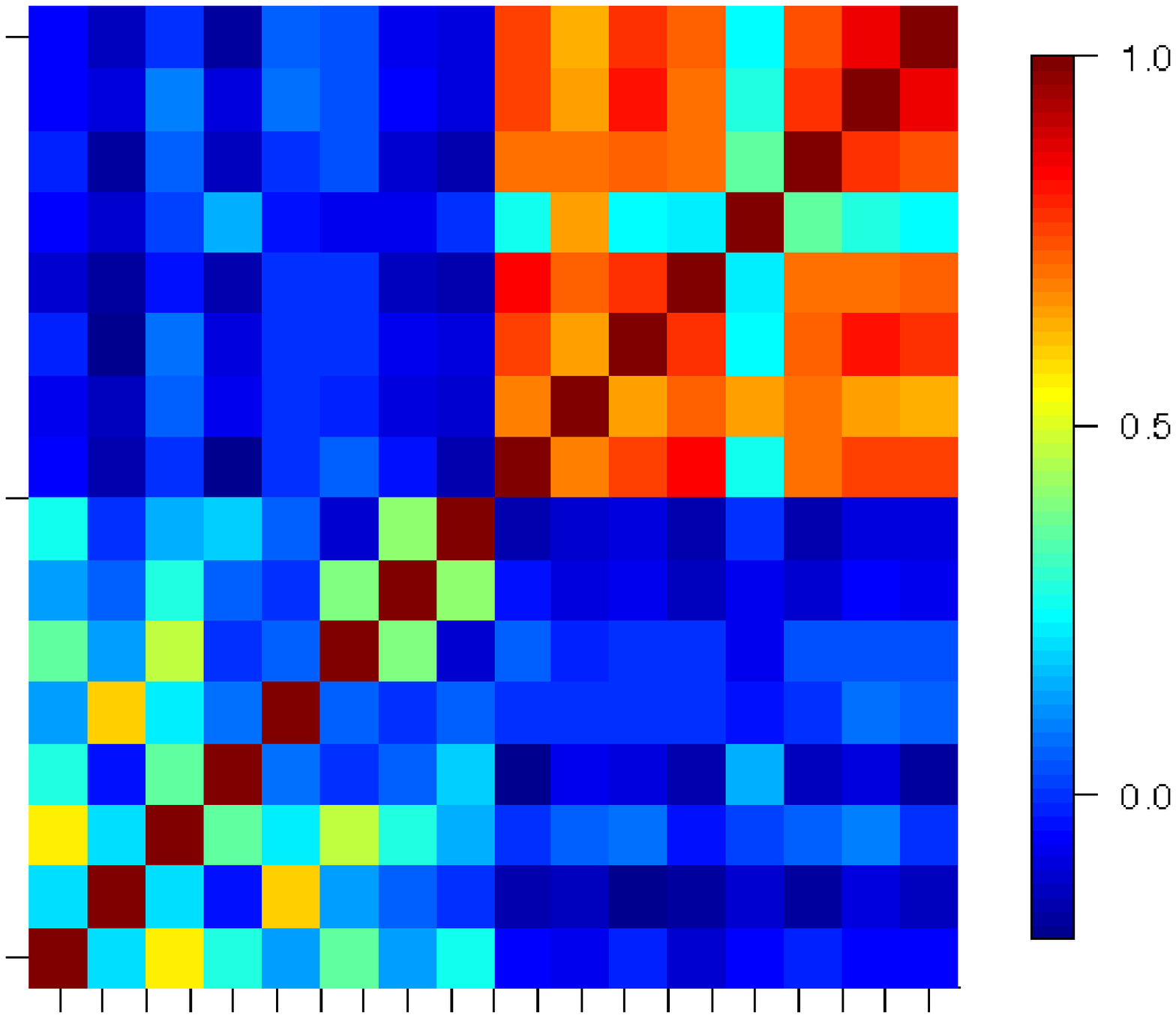} &
    \includegraphics[angle=-90,width=3in, scale=1]{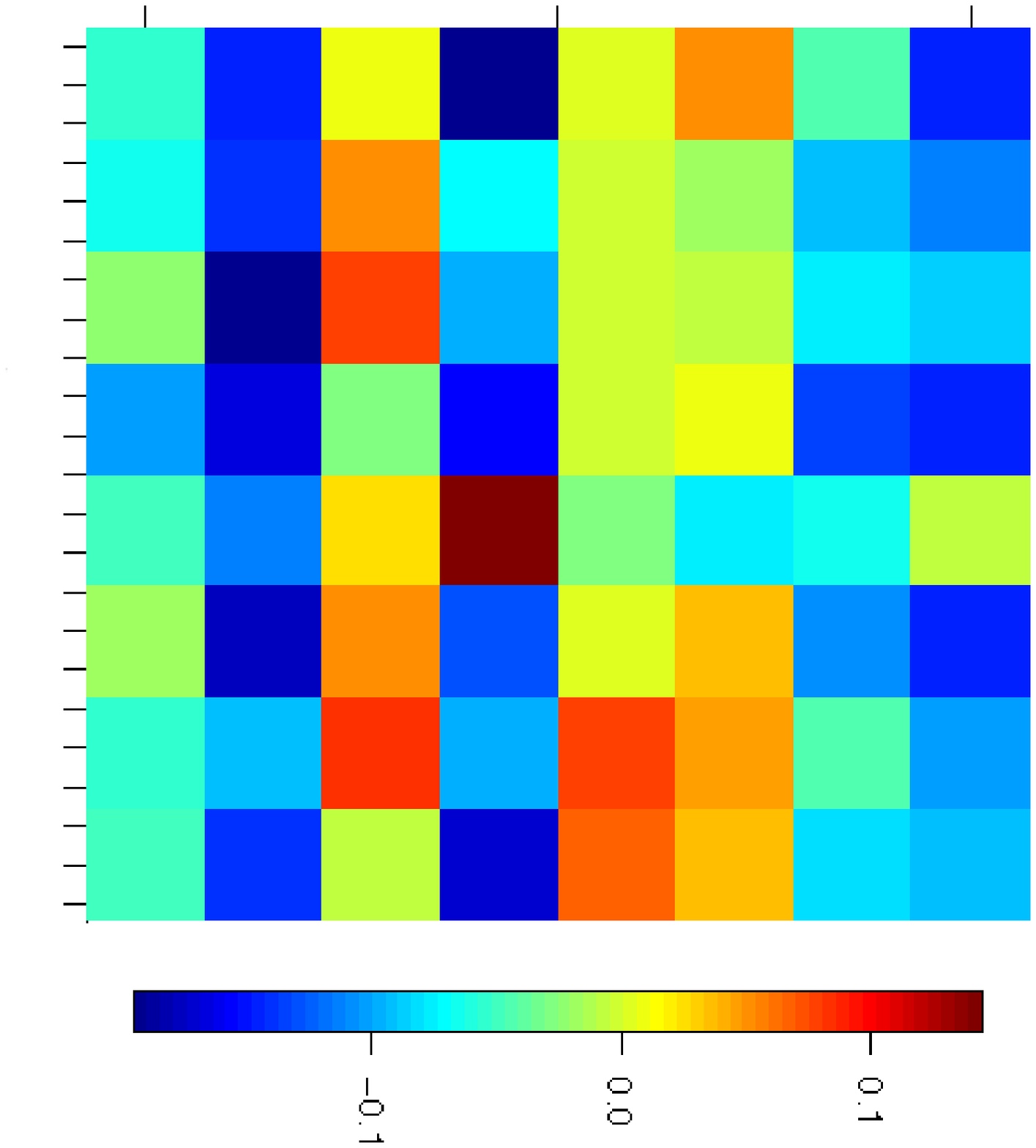}
\end{tabular}
\begin{minipage}{0.8\textwidth}
\bigskip
 \caption{\small Estimated covariance matrix between log(E2) and log(F2Iso) and the zoomed in cross covariance structure  by the shrinkage prior}
 \label{fig:E2F2Iso}
\end{minipage}
\end{figure}

Figures \ref{fig:E2F2Iso1} and \ref{fig:E2F2Iso} display the
estimated correlation structures for both within and across
responses for  the shrinkage and the MM priors respectively. The
left panels are the estimated correlation matrices between the two
responses and the right panels are the zoomed-in cross correlation
structures among the two responses. The left upper 8 by 8 matrix
(with the 8 responses from the cycles) is the correlation matrix for
response E2 across the cycle. The right lower 8 by 8 matrix is the
correlation matrix for the eight F2Iso responses across the cycle.
For example, the (2,3)rd cell is the correlation between the second
visit and the third visit of response E2; the (10,11)th cell is the
correlation between the second visit and the third visit of response
F2Iso. The upper right (or the lower left) 8 by 8 matrix is the
cross correlation among responses of E2 and F2Iso across the two
cycles. For example, the (1,9)th cell is the correlation between the
first visit of E2 and the first visit of F2Iso.

Estimated correlation structures through the MM priors
 help us have a better understanding of the relation
between estrogen levels and F2 Isoprostanes during the menstrual
cycle: finding more visits with stronger correlations between the
two responses. The analysis shows that the correlations appear to
differ slightly across the menstrual cycle, with the
cross-correlations being low in general.  Further, the 5th visit for
F2Iso is much less correlated than the others.  This could be due to
the fact that the mean values of F2Iso tend to be lowest at this
point during the cycle (around ovulation when estrogen levels are
high), but otherwise are not varying as much at the other visits.
Estimates from the shrinkage priors fail to pick up most of the
stronger correlations between visits.

\section{Discussion}
This article has proposed two new methods for Bayesian model
selection of fixed and random effects in continuous models. Our
approaches rely on  shrinkage priors and MM priors to the setting of
variable selection of multivariate, correlated random effects with
large dimension. Clear advantages over earlier approaches include
robustness, efficiency of posterior computation and overcoming the
order dependence problem.

Our proposed approach is advantageous in that fixed and random
effects are selected simultaneously. In particular, the prior and
computational algorithm represent a useful alternative to approaches
that rely on inverse-Wishart priors for variance components. There
is an increasing realization that inverse-Wishart priors are a poor
choice, particularly when limited prior information is available.
Although we have focused on LMEs of the Laird and Ware (1982) type,
it is straightforward to adapt our methods to a broader class of
linear mixed models, accommodating varying coefficient models,
spatially correlated data, and other applications.

\section{Acknowledgement}
This work was supported in part by the Intramural Research Program
of the \emph{Eunice Kennedy Shriver} National Institute of Child
Health and Human Development, National Institutes of Health.

\bibliographystyle{jasa}
\bibliography{ref}

\end{document}